\documentclass[12pt]{article}
\usepackage{epsf}
\usepackage{cite}
\usepackage{amssymb}
\def\g{\gamma}
\def\m{\mu^2}
\def\G{\Gamma}
\def\i{\prime}
\def\am{(\alpha_1-\beta_1)}
\def\ap{(\alpha_1+\beta_1)}
\def\amn{\am_{\pi^0}}
\def\amc{\am_{\pi^{\pm}}}
\def\apc{\ap_{\pi^{\pm}}}
\def\apn{\ap_{\pi^0}}
\def\bm{(\alpha_2-\beta_2)}
\def\bp{(\alpha_2+\beta_2)}
\def\bmn{\bm_{\pi^0}}
\def\bmc{\bm_{\pi^{\pm}}}
\def\bpc{\bp_{\pi^{\pm}}}
\def\bpn{\bp_{\pi^0}}
\def\gg{\g \g\to \pi^0 \pi^0}
\def\s{\sigma}
\def\f{\varphi}
\def\ggc{\g\g\to\pi^+\pi^-}
\def\gp{\g p\to\g\pi^+n}
\def\sig{\frac{d\s_{\gg}}{d\Omega}}

\def\tg{\theta^{*}}
\def\thq{\theta_{\g\g^\i}^{cm}}

\def\mpp{M_{++}}
\def\mm{M_{+-}}
\def\sp{s'}
\def\tp{t'}
\def\unit{10^{-4} {\rm fm}^3}
\def\unitq{10^{-4} {\rm fm}^5}
\def\be{\begin{equation}}
\def\ee{\end{equation}}
\def\beq{\begin{eqnarray}}
\def\eeq{\end{eqnarray}}
\begin{document}
\begin{center}
{\large\bf PION POLARIZABILITIES}\\
\vspace{0.5cm}

\underline{L.V. Fil'kov} and V.L. Kashevarov\\
{\it Lebedev Physical Institute, Leninsky Prospect 53, 119991 Moscow, Russia}
\end{center}
\section{Introduction}

Pion polarizabilities are fundamental structure parameters characterizing
the behavior of the pion in an external electromagnetic field.
The dipole and quadrupole polarizabilities are defined \cite{rad,fil2} through
the expansion of the non-Born helicity amplitudes of the Compton scattering
on the pion over $t$ at the fixed $s=\m$
\beq
\mpp(s=\m,t)&=&\pi\mu\left[2\am+\frac{t}{6}\bm\right]+{\cal O}(t^2),
\nonumber \\
\mm(s=\m,t)&=&\frac{\pi}{\mu}\left[2\ap+\frac{t}{6}\bp\right]+{\cal O}(t^2),
\label{mpm}
\eeq
where $s$ ($t$) is the square of the total energy (momentum transfer) in the
$\g\pi$ c.m. system and $\mu$ is the pion mass.
The dipole electric ($\alpha_1$) and magnetic ($\beta_1$) pion polarizabilities
measure the response of the pion to quasistatic electric and magnetic
fields. On the other hand, the parameters $\alpha_2$ and $\beta_2$
measure the electric and magnetic quadrupole moments induced in the pion
in the presence of an applied field gradient.
In the following the dipole and quadrupole polarizabilities are given in
units $\unit$ and $\unitq$, respectively.

The values of the pion polarizabilities are very sensitive to predictions of
different theoretical models. Therefore, an accurate experimental determination
of these parameters are very important for testing the validity of such
models.

By now the values of the pion polarizabilities were determined by analysing
processes $ \pi^-A\to \g\pi^-A$, $\gp$, and $\g\g\to\pi\pi$.
In the present work we mainly analyse the results obtained in recent works
\cite{fil2,mami,fil1,fil3}

\section{$\pi^0$ meson polarizabilities}

At present the most reliable method of a determination of the $\pi^0$ meson
polarizabilities is an analysis of the process $\gg$ in the energy region
up to $\sim 2$ GeV
where the cross section of this process is very sensitive
to the values of the $\pi^0$ polarizabilities. This process is described by
the following invariant variables:
\be
t=(k_1+k_2)^2, \quad  s=(q_1-k_1)^2, \quad  u=(q_1-k_2)^2,
\ee
where $q_1(q_2)$ and $k_1(k_2)$ are the pion and photon four-momenta.
The cross section of the process $\gg$ is expressed through the helicity
amplitudes as follows
\be
\sig=\frac{1}{256\pi^2}\sqrt{\frac{(t-4\m)}{t^3}}\left\{t^2 |\mpp|^2+
\frac{1}{16}t^2(t-4\m)^2 \sin^4\tg |\mm|^2\right\},
\ee
where $\tg$ is the angle between the photon and the pion in the c.m.s. of
the process under consideration.

In order to analyse the process $\gg$ we
constructed dispersion relations (DRs) at fixed $t$ with one subtraction at
$s=\m$ for the helicity amplitude $\mpp (s,t)$ \cite{fil2}
\beq
Re \mpp (s,t)=Re \mpp (s=\m,t)
&+&\frac{(s-\m)}{\pi}P~\int\limits_{4\m}^{\infty}d\sp~Im\mpp(\sp,t)\left[
\frac{1}{(\sp-s)(\sp-\m)}\right.  \nonumber \\
&-&\left.\frac{1}{(\sp-u)(\sp-\m+t)}\right].
\label{dr1}
\eeq
Via the cross symmetry these DRs are identical to DRs with two subtractions.
The subtraction function $Re \mpp (s=\m,t)$ was determined
with the help of the DRs at fixed $s=\m$ with two subtractions using
the cross symmetry between the $s$ and $u$ channels
\beq
&&Re \mpp(s=\m,t)= \mpp (s=\m,0)+\left.t\frac{d\mpp (s=\m,t)}{dt}
\right|_{t=0} \nonumber \\
&& +\frac{t^2}{\pi}\left\{P\int\limits_{4\m}^{\infty}
\frac{Im\mpp(\tp,s=\m)~d\tp}{\tp^2(\tp-t)} +\int\limits_{4\m}^{\infty}
\frac{Im\mpp(\sp,u=\m)~d\sp}{(\sp-\m)^2(\sp-\m+t)}\right\}.
\label{sub}
\eeq

The DRs for the amplitude $\mm(s,t)$ had the same expressions
(\ref {dr1}) and (\ref{sub})
with substitutions: $\mpp\to \mm$ and $Im\mpp \to Im\mm$.

The subtraction constants were expressed through the sum and the
difference of the electric and magnetic polarizabilities taking into
account Eq. (1):
\beq
&&\mpp(s=\m,t=0)=2\pi\mu\amn, \quad
\left.\frac{d\mpp(s=\m,t)}{dt}\right|_{t=0}=\frac{\pi\mu}{6}\bmn ,
\label{a-b} \\
&&\mm(s=\m,t=0)=\frac{2\pi}{\mu}\apn, \quad
\left.\frac{d\mm(s=\m,t)}{dt}\right|_{t=0}=\frac{\pi}{6\mu}\bpn .
\label{a+b}
\eeq
 These DRs were used to fit the experimental data to the total cross section
of the process $\gg$. The DRs were saturated by the contribution of
$\rho(770)$, $\omega(782)$, and $\phi(1020)$ mesons in the $s$ channel and
$\s$, $f_0(980)$, $f_0(1370)$, $f_2(1270)$, and $f_2(1525)$ in $t$ channel.
The polarizabilities $(\alpha_1\pm \beta_1)_{\pi^0}$ and
$(\alpha_2\pm \beta_2)_{\pi^0}$  and parameters of the $\s$ meson were
considered as free parameters.

For the reaction under consideration the Born term is equal to zero and the
main contribution in the energy region $\sqrt{t}=$270--825 MeV is
determined by $S$ wave. So, this process gives a good possibility to search
for the $\s$ meson.
The parameters of such a $\s$ meson and the values of the dipole
polarizabilities have been found from the fit to the
experimental data \cite{mars} in the energy regions 270--825 MeV
and 270--2000 MeV, respectively \cite{fil3}.
The values of the quadrupole polarizabilities have been determined from
the fit to experimental data \cite{mars,bien} in the energy region
270--2250 MeV \cite{fil2}.

The result of the fit to the experimental data for the total cross
section \cite{mars,bien} in the energy region from threshold to
2250 MeV is presented in Fig. 1 by solid curve \cite{fil2}.
\begin{figure}[h]
\epsfxsize=8cm   
\epsfysize=7.5cm 
\centerline{
\epsffile{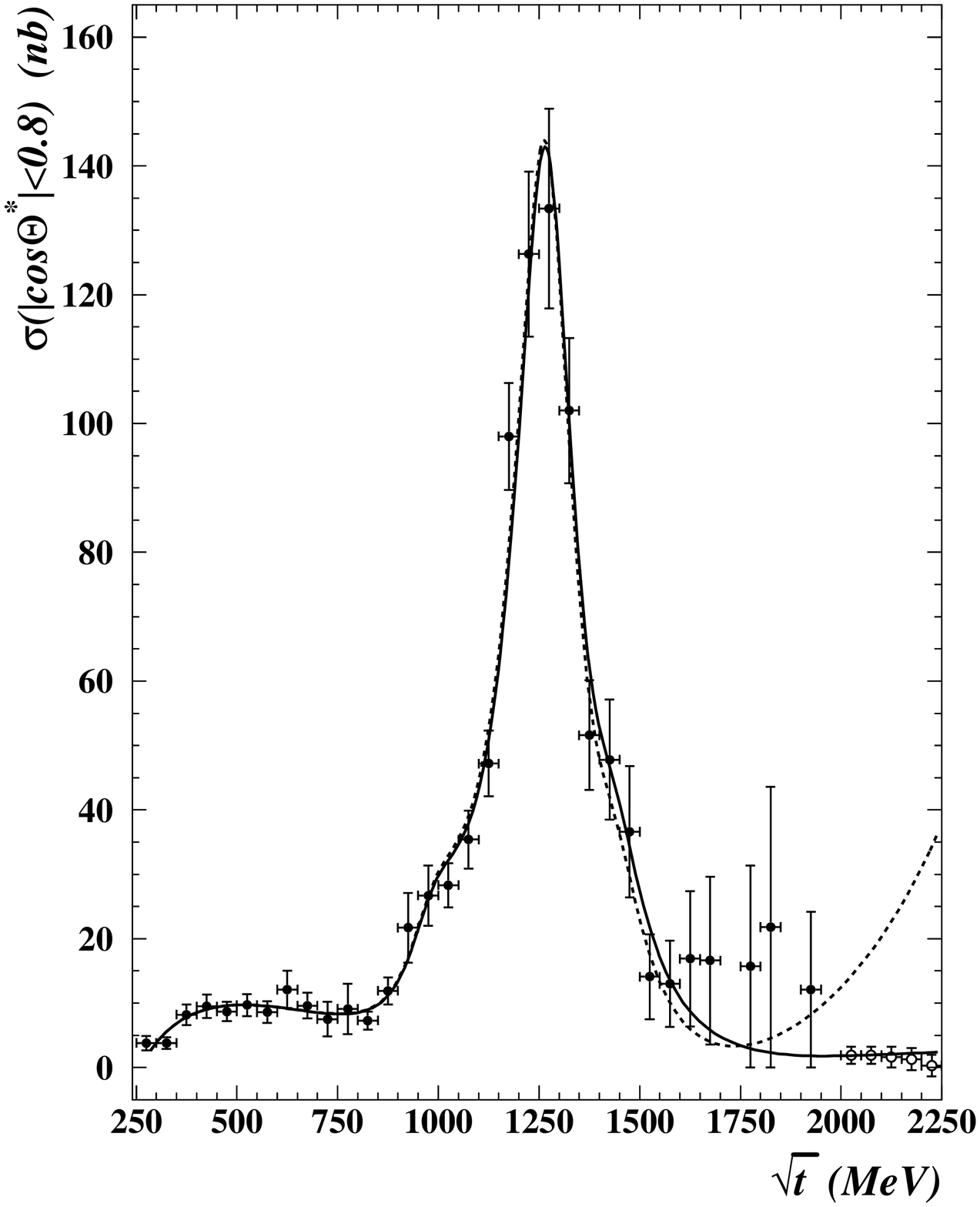} }    
\caption{The total cross section  of the reaction $\gg$.
}
\end{figure}
The dashed curve corresponds to the quadrupole polarizabilities calculated
with the help of the dispersion sum rules (DSRs) from Ref. \cite{fil2}.
The full circles are data from Ref. \cite{mars} and the open ones are
data from Ref. \cite{bien}.

The following parameters of the $\s$ meson have been determined in the Ref.
\cite{fil3}:
$m_{\s}=547\pm 45$~MeV, $\G_\s=1204\pm 362$ MeV,
$\G_{\s\to\g\g}=0.62\pm 0.19$~keV.
This value of $\G_{\s\to\g\g}$ differs strongly from the result of Ref.
\cite{penn} [$\G_{\s\to\g\g}=(3.8\pm 1.5)$~keV].
It should be noted that the use of the value of
$\G_{\s\to\g\g}=(3.8\pm 1.5)$~keV  in the analysis \cite{fil2} leads
to a strong deviation from the experimental data on the total cross section
of the process under consideration.

The values of the dipole and quadrupole polarizabilities found in the fits
\cite{fil2,fil3}
are listed in Table 1 together with results obtained in Refs. \cite{kal,ser}
and prediction of DSRs \cite{fil2} and two loop calculations in the frame
of ChPT \cite{bell,ivan1}.
\begin{table}
\caption{The dipole and quadrupole polarizabilities of the $\pi^0$ meson}
\centering
\begin{tabular}{|c|c|c|c|}\hline
        &fit                        &DSRs \cite{fil2} &ChPT  \\ \hline
$\amn$   &$-1.6\pm 2.2$ \cite{fil3} &$-3.49\pm 2.13$ &$-1.9\pm 0.2$
\cite{bell}  \\ 
        &$-0.6\pm 1.8$ \cite{kal}   &                & \\ \hline
$\apn$   &$0.98\pm 0.03$ \cite{fil3}&$0.802\pm 0.035$&$1.1\pm 0.3$
\cite{bell}  \\ 
        &$1.00\pm 0.05$ \cite{ser} &                & \\ \hline
$\bmn$   &$39.70\pm 0.02$\cite{fil2}&$39.72\pm 8.01$ &$37.6\pm 3.3$
\cite{ivan1}       \\ \hline
$\bpn$   &$-0.181\pm 0.004$\cite{fil2}&$-0.171\pm 0.067$& 0.04
\cite{ivan1}       \\ \hline
\end{tabular}
\end{table}
The obtained values of the
sum and difference of the dipole polarizabilities of $\pi^0$ and the
difference of its quadrupole polarizabilities do not conflict within the
errors with the predictions  of DSRs and ChPT.
However, there are very big errors in the
experimental values for the difference of the dipole polarizabilities.
Therefore, it is difficult to do a more unambiguous conclusion. As for
the sum of the quadrupole polarizabilities of $\pi^0$, the DSR result
agrees well with the experimental value, but the ChPT predicts a positive
value in contrast to experimental result. However, as it was noted in
Ref. \cite{ivan1}, this quantity was obtained in  a two-loop approximation,
which is a leading  order result for this sum, and one expects substantial
corrections to it from three-loop calculations.

It should be noted that the values of the difference and the sum of the
quadrupole polarizabilities found from the fit have very small
errors. They are the errors of the fitting.
This is a result of a very high sensitivity of the total cross
section of the process $\gg$ at $\sqrt{t}>1500$ MeV to values of these
parameters. In order to estimate real values of errors of these difference
and sum, model errors should be added.

\section{Measurement of the $\pi^+$ meson polarizabilities via the
$\gp$ reaction}

The pion polarizability can be extracted from experimental data on
radiative pion photoproduction, either by  extrapolating these data to the
pion pole \cite{drec,fild,filp,walch},
or by comparing the experimental cross section with the predictions of
different theoretical models.
The extrapolation method was first suggested in \cite{chew} and has been widely
used for the determination of cross sections and phase shifts of
elastic
$\pi\pi$-scattering from the reaction \mbox{$\pi N\rightarrow \pi \pi N$}.
For investigations of $\g\pi^+$-scattering this method was first used in
\cite{ayb,lebed}.

However, in order to obtain a reliable value of the pion
polarizability, it is necessary to obtain the experimental data on pion
radiative photoproduction with small errors over a sufficiently
wide region of $t$, in particular very close to $t=0$ \cite{ahr,ahr1}.

It should be noted that there is an essential difference in
extrapolating the data of the processes $\pi N\to \pi\pi N$ and
$\g p\to \g\pi N$. In the former case, the pion pole
amplitude gives the main contribution in a certain energy region. This
permits one to constrain the extrapolation function to be zero
at $t=0$ providing a precise determination of the amplitude. In the
case of radiative pion photoproduction, the pion pole amplitude
alone is not gauge invariant and we must take into account all pion and
nucleon pole amplitudes. However, the sum of these amplitudes does not
vanish at $t=0$. This complicates
the extrapolation procedure by increasing the requirements on the
accuracy of the experimental data.

As the accuracy of the data \cite{mami} was not sufficient for a reliable
extrapolation, the values of the pion polarizabilities have been obtained
from a fit of the cross section calculated by different theoretical
models to the data.

The theoretical calculations of the cross section for the reaction
$\gp$ showed that the contribution of
nucleon resonances is suppressed for photons
scattered backward in the c.m.s. of the reaction $\g\pi\to\g\pi$.
Moreover, integration over $\f$ and $\thq$ essentially decreases the
contribution of resonances from the crossed channels.
On the other hand, the difference \mbox{$\am_{\pi^+}$} gives the biggest
contribution to the cross section
for $\thq$ in the region of $140^{\circ}-180^{\circ}$.
Therefore, we considered the cross section of radiative pion
photoproduction integrated over $\f$ from $0^{\circ}$ to $360^{\circ}$
and over $\thq$ from $140^{\circ}$ to $180^{\circ}$,
\be
\int_0^{360^{\circ}}d\varphi\int_{-1}^{-0.766}d\cos\thq\;
\frac{d\s_{\gp}}{dtds_1d\Omega_{\g\g}},
\ee
where $s_1=(q_1+k_1)^2$ is the square of the total energy in c.m. system
for the $\g\pi\to\g\pi$ reaction, $t=(p_p-p_n)^2\simeq -2mT_n$
is the square of the
momentum transfer for the $\gp$ reaction and $T_n$ is the kinetic energy of
the neutron.

The cross section of the process $\gp$ has been
calculated in the framework of two different models.
In the first model (model-1) the contribution of all
the pion and nucleon pole diagrams was taken into account
using pseudoscalar pion-nucleon coupling \cite{unk}.

The second model (model-2) included the nucleon and the
pion pole diagrams without the anomalous magnetic moments of the nucleons,
and in addition
the contributions of the resonances $\Delta (1232)$, $P_{11}(1440)$,
$D_{13}(1520)$, $S_{11}(1535)$, and $\s$ meson.

The experiment on the radiative $\pi^+$ meson photoproduction was carrie out
at the Mainz Microtron MAMI in the energy region 537 MeV$<E_\g<$817 MeV.

To increase our confidence that
the model dependence of the result was under control we limited ourselves
to kinematic regions where the difference between
model-1 and model-2 did not exceed $3\%$ when $\am_{\pi^+}$ was
constrained to zero.
First, the kinematic region, where the contribution of the
pion polarizability is negligible, i.e. the region
$1.5\m\le s_1<5\m$, was considered.

In Fig.~2, the experimental data for the differential
\begin{figure}[h]
\begin{minipage}{0.46\textwidth}
\epsfxsize=\textwidth
\epsffile{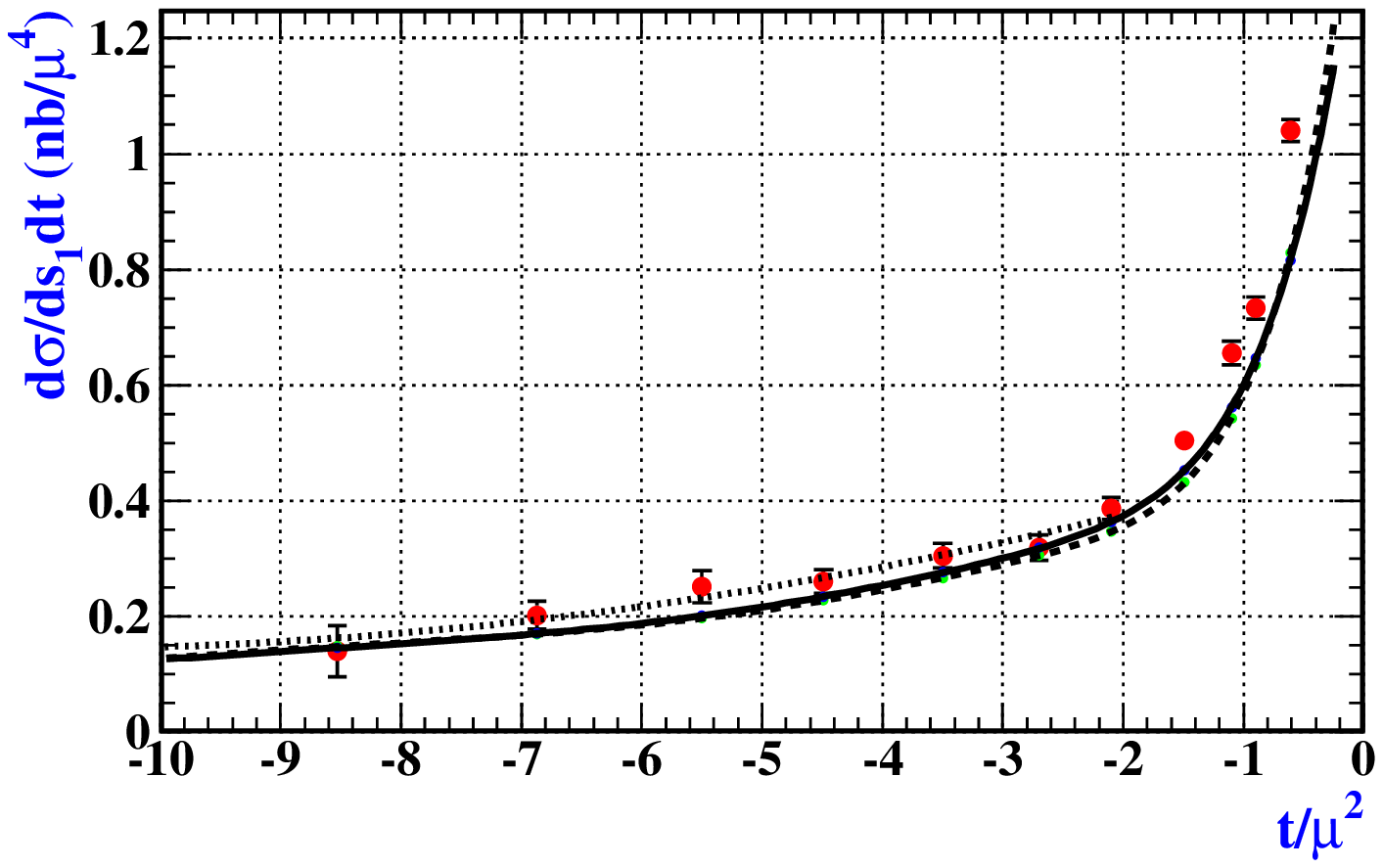}
\caption{The differential cross section of the process $\gp$ averaged
over the full photon beam energy interval and over $s_1$ from 1.5$\m$ to
5$\m$}
\end{minipage}
\qquad
\begin{minipage}{0.46\textwidth}
\epsfxsize=\textwidth
\epsffile{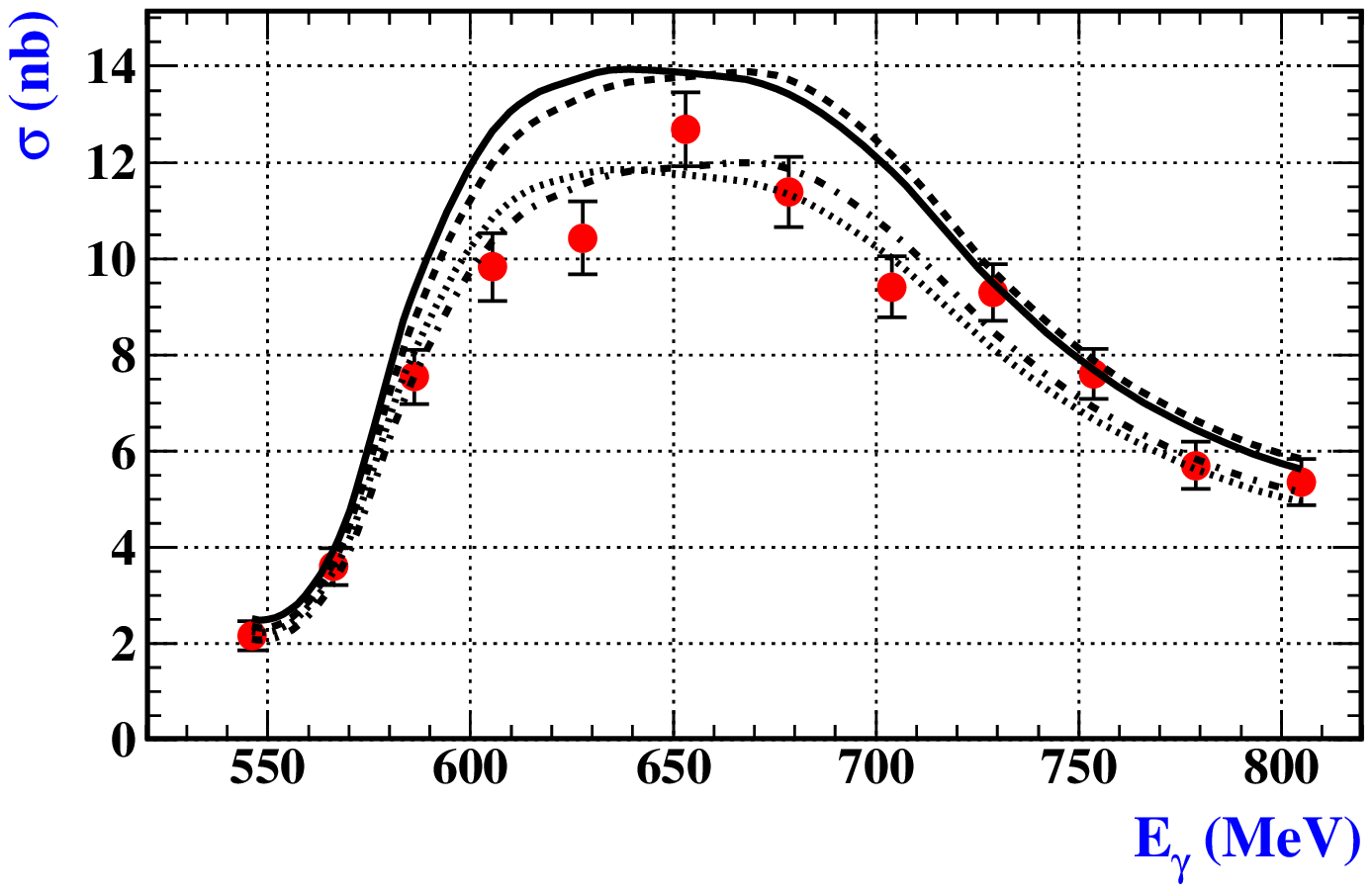}
\caption{The differential cross section of the process $\gp$ integrated
over $s_1$ and $t$ in the region, where the contribution of the  pion
polarizability is biggest.}
\end{minipage}
\end{figure}
cross section, averaged over the full photon beam energy interval from
537 MeV up to 817 MeV and over $s_1$ in the indicated interval,
are compared to predictions of model-1 (dashed
curve) and model-2 (solid curve). The dotted
curve is the fit of the experimental data in the region of
$-10\m<t<-2\m$.
As seen from this figure, the theoretical curves are very close to the
experimental data.
The small difference between the theoretical curves and the experimental
data was used for a normalization of the experimental data.

Then we investigated the kinematic region where the polarizability
contribution is biggest. This is the region $5\m\le s_1<15\m$ and
$-12\m<t<-2\m$.
In the range $t>-2\m$ the polarizability contribution is small
and also the efficiency of the TOF is not well known here. Therefore,
we have excluded this region.
Fig. 3 presents the cross section of the process $\gp$
integrated over $s_1$ and $t$ in the region where the contribution of the
pion polarizabilities is biggest end the difference between of the theoretical
models does not exceed 3\%. The dashed and dashed-dotted lines are the
predictions of model-1 and the solid and dotted lines of model-2 for
$\am_{\pi^+}=0$ and 14, respectively. As a result, the following value of the
difference of the charged pion dipole  polarizabilities has been obtained:
\be
\am_{\pi^+}=11.6\pm 1.5_{stat}\pm 3.0_{syst}\pm 0.5_{mod}.
\ee\label{mami}

\section{Determination of the charged pion polarizabilities\\ from the process
$\ggc$}

Attempts to determine the charge pion dipole polarizabilities from the
reaction $\ggc$ suffered greatly from theoretical and experimental
uncertainties. The analyses \cite{bab,holst,kal} have been performed in the
region of the low energy ($\sqrt{t}<700$~MeV,
where $t$ is the square of the total energy in $\g\g$ c.m.system).
In this region values of the experimental
cross sections of the process under consideration \cite{pluto,dm1,dm2,mark}
are very ambiguous. As a result, the values of $\alpha_{1\pi^{\pm}}$ found
lie in the interval 2.2--26.3. The analyses of the data of Mark II
\cite{mark} have given $\alpha_{1\pi^{\pm}}$ close to ChPT result. However,
even changes of the dipole polarizabilities by more than 100\% are still
compatible with the present error bars in the energy region considered
\cite{holst}.

In the work \cite{fil1} we constructed the DRs similar to those of
Ref. \cite{fil2}
for the amplitudes of the process $\ggc$. But in this case the Born term
does not equal to 0. Using the DRs allows one to avoid
the problem of double counting and the subtractions in the DRs provide
a good convergence of the integrand expressions of these DRs and, therefore,
increases the reliability of the calculations.
The DRs for the charged pions are saturated  by the contributions of
the $\rho(770)$, $b_1(1235)$, $a_1(1260)$, and $a_2(1320)$ mesons in
the $s$ channel and $\s$, $f_0(980)$, $f_0(1370)$, $f_2(1270)$, and
$f_2(1525)$ in the $t$ channel.

These DRs, where the charged pion dipole and quadrupole polarizabilities were
free parameters, were used to fit to the experimental data for the total
cross section
\cite{tpc,mark,cello,venus,aleph,belle}
in energy region from threshold to 2500 MeV.
The best result of this fit is presented in Fig. 4 by the solid line.
\begin{figure}[h]
\epsfxsize=15cm       
\epsfysize=9cm 
\centerline{
\epsffile{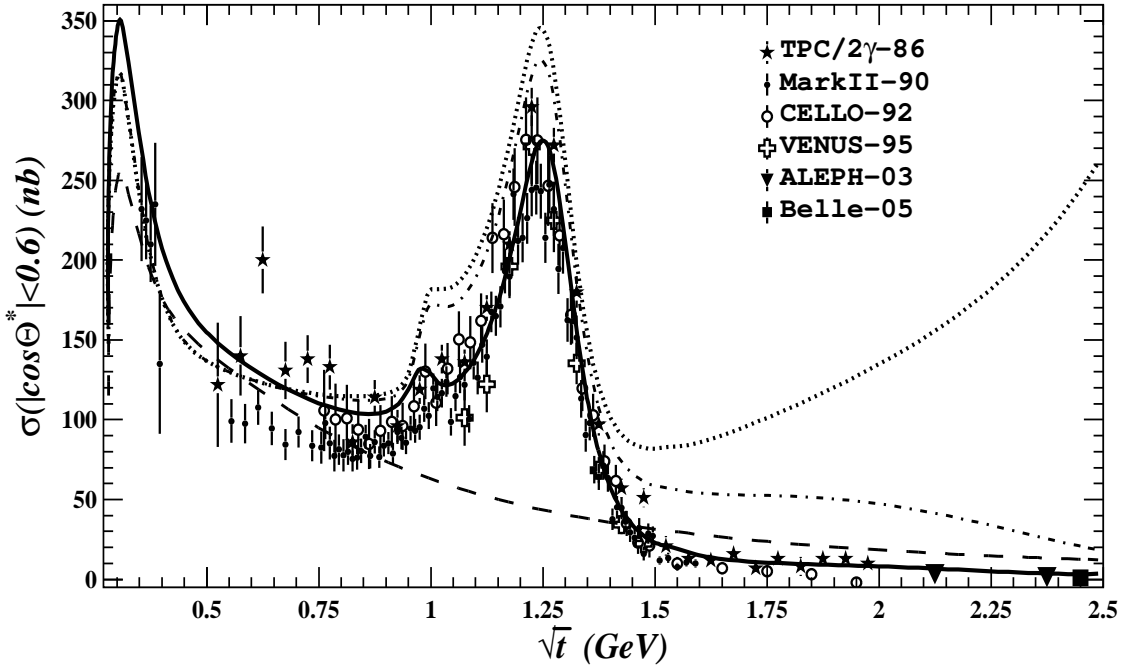}}
\caption{The total cross section of for the reaction $\ggc$
(with $|\cos{\theta^*}|<0.6$).}
\end{figure}
This solid curve well describes the experimental data in whole energy region
under investigation.
As a result, we have found the dipole polarizabilities of the charged
pions and determined their quadrupole polarizabilities for the first time.
The values of the polarizabilities found in the work \cite{fil1} and
the predictions of DSRs
\cite{fil2} and two-loop ChPT \cite{ivan2} are listed in Table~2.
\begin{table}
\caption{The dipole and quadrupole polarizabilities of the charged pions.}
\centering
\begin{tabular}{|c|l|l|c|c|} \hline
         &                       &                &\multicolumn{2}{|c|}{
ChPT \cite{ivan2}} \\ \cline{4-5}
         &\qquad fit \cite{fil1} &DSRs \cite{fil2} &to one-loop & to two-loops
\\ \hline
$\amc$   &$13.0^{+2.6}_{-1.9}$   &$13.60\pm 2.15$ &6.0 &5.7 [5.5]  \\ \hline
$\apc$   &$0.18^{+0.11}_{-0.02}$ &$0.166\pm 0.024$&0 & 0.16 [0.16] \\ \hline
$\bmc$   &$25.0^{+0.8}_{-0.3}$   &$25.75\pm 7.03$ &11.9&16.2 [21.6] \\ \hline
$\bpc$   &$0.133\pm 0.015$       &$0.121\pm 0.064$&0&-0.001 [-0.001]
\\ \hline
\end{tabular}
\end{table}
The numbers in brackets correspond to the order $p^6$ low energy constants
from Ref. \cite{nc}.
As seen from this Table, the all values of polarizabilities found in Ref.
\cite{fil1} are in good
agreement with the DSR predictions \cite{fil2}.

The dashed curve in Fig. 4 is the Born term contribution.
The dotted curve is a result of calculations using the DRs
when $\bmc$ and $\bpc$ equal to the respective values in Table~2.
but the dipole polarizabilities are taken from ChPT calculations \cite{burgi}
as $\amc=4.4$ and $\apc=0.3$.
The dashed-dotted curve presents a result of the fit to the experimental data
when the quadrupole polarizabilities are the free parameters and the values
of the dipole  polarizabilities are fixed by ChPT calculations \cite{burgi}.
The both last curves are close to calculations in Ref. \cite{holst}
in the energy region up to 700 MeV, however they differ strongly from all
experimental data on the total cross section at higher energies.

The fits of the data to the total cross section for the separate works
\cite{mark,tpc,cello,venus} were used to estimate the errors of the values of
charged pion polarizabilities found.

The angular distributions of the differential cross section of the
process $\ggc$ at different energies are shown in Fig. 5.
\begin{figure}[h]
\epsfxsize=8.6cm
\epsfysize=10.5cm
\centerline{
\epsffile{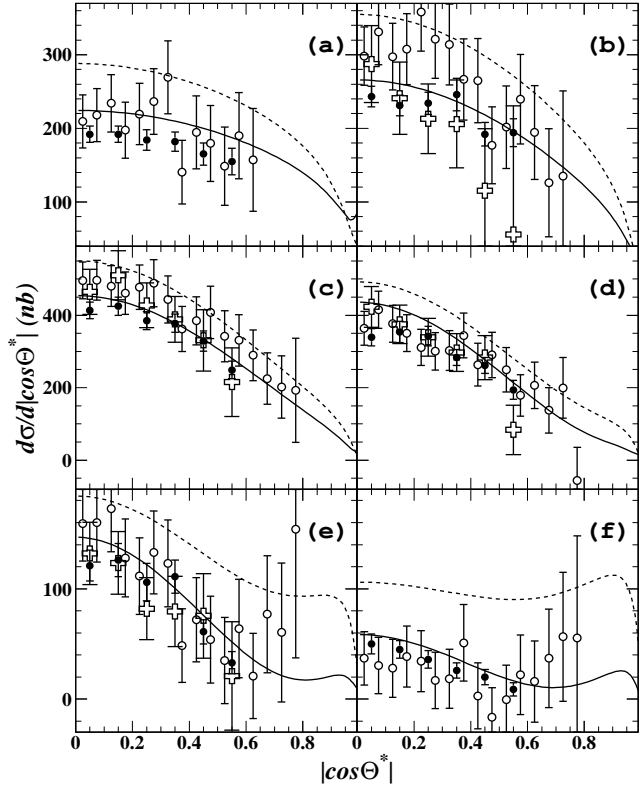}}
\caption{Angular distributions of the differential cross sections for the
following energy intervals: (a) -- 0.95--1.05 GeV, (b) -- 1.05--1.15 GeV,
(c) -- 1.15--1.25 GeV, (d) -- 1.25--1.35 GeV, (e) -- 1.35--1.45 GeV,
(f) -- 1.45--1.55 GeV.
The designations of the experimental data are the same
as in Fig. 4.}
\end{figure}
The solid and dashed curves are the results of calculations
using our and the ChPT fits (the latter when the values of the dipole
polarizabilities are fixed by ChPT \cite{burgi})
to the total cross sections in Fig. 4, respectively.
This figure demonstrates a good description of the angular distributions by the
solid curves with the polarizability values found
in the present work.
On the other hand, the calculations with the dipole polarizabilities from
ChPT \cite{burgi} contradict these experimental data, particularly at
higher energies.

\section{ Discussion}   

The experimental information available so far for the difference of the
dipole polarizabilities of charged pions is summarized in Table 3.
\begin{table}
\caption{The experimental data on $\amc$. In \cite{antip,holst,bab},
$\amc$ was determined by using a constraint
$\alpha_{1\pi^\pm}=-\beta_{1\pi^\pm}$.}
\centering
\begin{tabular}{|ll|c|} \hline
\multicolumn{2}{|l|}{Experiments} & $\amc$  \\ \hline
\multicolumn{2}{|l|}
{L.V. Fil'kov, V.L. Kashevarov (2006) \cite{fil1}}&
$13.0^{+2.6}_{-1.9}$ \\
$\ggc$: & MARK II \cite{mark}, &                 \\
\multicolumn{2}{|l|}{TPC/2$g$ \cite{tpc}, CELLO \cite{cello},} &  \\
\multicolumn{2}{|l|}{VENUS \cite{venus}, ALEPH \cite{aleph}, BELLE \cite{belle}}
&  \\ \hline
\multicolumn{2}{|l|}{$\gp$: MAMI (2005) \cite{mami}} &
$11.6\pm 1.5_{stat}\pm 3.0_{syst}\pm 0.5_{mod}$ \\ \hline
\multicolumn{2}{|l|}{A.E. Kaloshin, V.V. Serebryakov (1994) \cite{kal}}&
$5.25\pm 0.95$ \\
$\ggc$: &  MARK II \cite{mark} &     \\ \hline
\multicolumn{2}{|l|}{J.F. Donoghue, B.R. Holstein (1993) \cite{holst}} & 5.4 \\
$\ggc$:   & Mark II \cite{mark}    &   \\ \hline
\multicolumn{2}{|l|}{D. Babusci {\em et al.} (1992) \cite{bab}} &     \\
$\ggc$:   & PLUTO  \cite{pluto} & $38.2\pm 9.6\pm 11.4$  \\
    & DM 1  \cite{dm1} & $34.4\pm 9.2$      \\
    & DM 2  \cite{dm2} & $52.6\pm 14.8$      \\
    & Mark II \cite{mark} & $4.4\pm 3.2$     \\ \hline
\multicolumn{2}{|l|}{$\gp$: Lebedev Phys.Inst. (1984) \cite{ayb}} &
$40\pm 24$      \\  \hline
\multicolumn{2}{|l|}{$\pi^{-}Z\rightarrow\g \pi^{-} Z$: Serpukhov (1983)
\cite{antip}} & $13.6\pm 2.8\pm 2.4$     \\  \hline
\end{tabular}
\end{table}

The difference of the dipole polarizabilities of charged pions found
from the analysis of the process $\ggc$  \cite{fil1}
agrees very well with results obtained from the scattering of high
energy $\pi^-$ mesons off the Coulomb field of heavy nuclei \cite{antip}
and from the radiative photoproduction of $\pi^+$ from the proton at MAMI
\cite{mami}  and in Lebedev Physical Institute \cite{ayb} (see Table~3).
However, these values of $\amc$  deviate substantially
from the calculations in the framework of ChPT \cite{ivan2,burgi}.

The analyses \cite{kal} and \cite{holst} of the data of MARK II \cite{mark}
have given the values of $\amc$ close to the ChPT result.
However, they have been determined in the energy region $\sqrt{t}<700$ MeV,
were these data have big errors. Moreover, these small values lead to the
strong deviation from the experimental data at higher energies (see Fig. 4).

One of the possible reasons for the small value of $\amc$ predicted by ChPT
could be the neglect of the contribution of the $\s$ meson. As has been
shown in Ref. \cite{fil2}, this resonance gives the main contribution to
DSRs for $\amc$.

The difference of the quadrupole polarizabilities $\bmc$ (see Table 2)
disagrees with the present two-loop ChPT calculations \cite{ivan2}.
One of the sources of such a disagreement is a poor knowledge of low energy
constants. Moreover, it should be noted that in this case the two-loop
calculation generates nearly a 100\% contribution as compared to one-loop
result.

Calculations of $(\alpha_{1,2}+\beta_{1,2})$ at order $p^6$ determine only the
leading order term in ChPT. Therefore, contributions at $p^8$ could be
essential, and considerably more work required to put the ChPT prediction
on a firm basis \cite{ivan2}.

\section{Summary}

We have reviewed and analysed the data on the pion polarizabilities obtained.

1. The values of the dipole and quadrupole polarizabilities of $\pi^0$
have been obtained from the fit of the experimental data \cite{mars,bien} to
the total cross section of the process $\gg$ in the energy region from
threshold to 2250 MeV.
The values of $(\alpha_1\pm \beta_1)_{\pi^0}$ and
$\bmn$ do not conflict within the errors with the ChPT prediction. However,
two-loop ChPT calculations have given opposite sign for the $\bpn$.

2. The value of $\amc$ found in Ref. \cite{fil1} from the fit of
all available at present experimental data to the total cross section (with
$|\cos{\theta^*}|<0.6$) of the process $\ggc$ in the energy region from
threshold to 2500 MeV is consisted with the results obtained at
MAMI (2005) ($\gp$), in Serpukhov (1983) ($\pi^-Z\to\g\pi^-Z$) and Lebedev
Physical Institute (1984) ($\gp$). However, all these results are at variance
with the ChPT calculations.

3. The values of the quadrupole polarizabilities
$(\alpha_2\pm \beta_2)_{\pi^\pm}$ found disagree with the present two-loop
ChPT calculations.

4.All values of the pion polarizabilities found in Refs.
\cite{fil1,fil2,fil3, mami,ayb,antip} agree with DSR predictions.
\vspace{0.2cm}

This research is part of the EU integrated initiative hadron physics project
under contract number RII3-CT-2004-506078 and was supported in part by the
Russian Foundation for Basic Research (Grant No. 05-02-04014).

\end{document}